# Task-Projected Hyperdimensional Computing for Multi-Task Learning

Cheng-Yang Chang[✉], Yu-Chuan Chuang, and An-Yeu (Andy) Wu

Graduate Institute of Electronics Engineering, National Taiwan University,
Taipei, Taiwan
{kevin, frankchuang}@access.ee.ntu.edu.tw, andywu@ntu.edu.tw

**Abstract.** Brain-inspired Hyperdimensional (HD) computing is an emerging technique for cognitive tasks in the field of low-power design. As an energy-efficient and fast learning computational paradigm, HD computing has shown great success in many real-world applications. However, an HD model incrementally trained on multiple tasks suffers from the negative impacts of catastrophic forgetting. The model forgets the knowledge learned from previous tasks and only focuses on the current one. To the best of our knowledge, no study has been conducted to investigate the feasibility of applying multi-task learning to HD computing. In this paper, we propose Task-Projected Hyperdimensional Computing (TP-HDC) to make the HD model simultaneously support multiple tasks by exploiting the redundant dimensionality in the hyperspace. To mitigate the interferences between different tasks, we project each task into a separate subspace for learning. Compared with the baseline method, our approach efficiently utilizes the unused capacity in the hyperspace and shows a 12.8% improvement in averaged accuracy with negligible memory overhead.

## 1 Introduction

In the era of IoT, edge computing with energy-efficient machine learning models keeps data processing close to end-users. This brings out numerous advantages, including lower latency, user security, and cost savings [1]. Meanwhile, multi-task learning (MTL) is grabbing attention recently since a single model can accommodate multiple cognitive tasks is more desirable for the future of IoT [2].

Brain-inspired Hyperdimensional (HD) computing emulates the operations of brains and handles cognitive tasks in a hyperdimensional space with well-defined vector space operations [3]. As an energy-efficient and fast-learning computational paradigm, HD computing has shown successful progress in many real-world applications such as gesture recognition [4], language recognition [5], and general bio-signal processing [6][7]. Moreover, HD computing can operate at an ultra low-power condition with lower latency through massively parallel bitwise operation [8]. These advantages make HD computing suitable for efficient signal processing, e.g., 2× lower energy at iso-accuracy when compared to a highly-optimized SVM on an ARM Cortex M4 [9]. However, an HD model incrementally trained on multiple tasks forgets the knowledge learned from previous tasks and only focuses on the current one. The phenomenon is called catastrophic forgetting [10]. To the best of our knowledge, no study has been conducted to overcome this problem and investigate the feasibility of applying MTL to HD computing.

This paper aims to establish a reliable MTL framework based on HD computing to minimize the negative impact of catastrophic forgetting. Over-parameterization in DNN implies that only a small subspace spanned by the optimal parameters is occupied by a given task [11]. Based on this phenomenon, [12] exploits the redundant subspace in DNN to superimpose multiple models into one. We are inspired by this concept and propose to exploit the unused capacity in the hyperspace to project each task into a separate subspace for learning. Our approach efficiently mitigates the interferences between different tasks and keeps the knowledge learned from numerous tasks stored in one HD model with minimal accuracy degradation.

The rest of the paper is organized as follows. Section 2 provides a review of HD computing. Section 3 describes the proposed Task-projected Hyperdimensional Computing (TP-HDC) for multi-task learning. Section 4 shows our experiment setting and simulation results. Finally, we conclude this paper in Section 5.

## 2  Review of Hyperdimensional Computing

HD computing is based on high-dimensional and dense binary vectors, called HD vectors. The components of HD vectors are binary with equally probable (-1)s and 1s. The processing flow chart of a general HD computing is shown in Fig. 1 and can be divided into the following four stages:

**Nonlinear Mapping to Hyperspace:** The main goal of mapping is to project a feature vector $x$ to HD vectors with dimensionality ($d$), where $x \in R^m$ with m components. Feature identifier (ID) is regarded as a basic field, and the actual value of the feature is the filler of the field. HD computing starts by constructing Item Memory ($IM$) and Continuous item Memory ($CiM$). $IM = \{ID_1, ID_2, ..., ID_m\}$, where $ID_k \in (-1,1)^d, k \in \{1,2,...m\}$ corresponds to the ID of the $k^{th}$ feature component. When $d$ is large enough, any two different HD vectors in $IM$ are nearly orthogonal, implying that $Cos(ID_i, ID_j) \cong 0, Ham(ID_i, ID_j) \cong 0.5, if\ i \neq j$ [13]. $Cos(\cdot)$ and $Ham(\cdot)$ are cosine similarity metric and normalized Hamming distance between the two vectors, respectively.

Continuous item memory (CiM) serves as the look-up table for the actual value of a feature. The procedure of establishing CiM first finds the maximum value and minimum value of each feature denoted as $V_{max}$ and $V_{min}$. The range between $V_{max}$ and $V_{min}$ is quantized to $\ell$ levels, and then an HD vector $L_1 \in (-1,1)^d$ is assigned to $V_{max}$, and $L_\ell \in (-1,1)^d$ is assigned to $V_{min}$. The HD model determines $L_1$ and $L_\ell$ at random, making them approximately orthogonal. $CiM = \{L_1, L_2, \ldots, L_\ell\}$, where $L_k \in (-1,1)^d, k \in \{1,2,\ldots \ell\}$, and every vector in $CiM$ corresponds to a range of actual value. The spatial relation of levels is preserved through adjusting the Hamming distance between $L_i$ and $L_j$ according to the difference of value to which the two HD vectors correspond. In other words, each value of the specific feature component will be associated with a vector proportionate to $L_1$ and $L_\ell$. Mapping of each feature value to hyperspace comprises quantizing and looking up the corresponding vectors $\{S_1, S_2, \ldots, S_m\}$ in $CiM$. After mapping each feature component of data to HD vectors, a set of two-vector pairs $I = \{(ID_1, S_1), (ID_2, S_2), \ldots, (ID_m, S_m)\}$ can readily be used in the next stage with the vector space operations.

**Encoding:** The HD model conducts the binding operation, bitwise XOR operation ($\oplus$) between two HD vectors, for each two-vector pair in $I$. After that, the resulting $m$ HD vectors in the set $I$ is accumulated by the bundling operation, bitwise addition (+) between HD vectors. Followed by binarization with sign function denoted as $[\cdot]$, data can be encoded as (1) and represented by the resulting binary HD vector $T \in (-1,1)^d$.

$$T = \sum_{i=1}^{m} ID_i \oplus S_i = [ID_1 \oplus S_1 + ID_2 \oplus S_2 + \cdots + ID_m \oplus S_m] \quad (1)$$

**Training:** All training samples go through the previous two stages and the resulting vector $T$ is sent to the associative memory (AM) for training. Training samples of the same class denoted as $T_i$ for the $i^{th}$ class are bundled together to form a class HD vector, as shown in (2). $n_i$ means the number of training samples of the $i^{th}$ class. For a $k$-class classification task, AM comprises $k$ class HD vectors, denoted as $\{C_1, C_2, \ldots, C_k\}$.

$$C_i = \sum_{j} T_i^j = [T_i^1 + T_i^2 + \cdots + T_i^{n_i}] \quad (2)$$

**Classification:** In the inference phase, an unseen testing data would go through the same processing flow of mapping and encoding in the training phase and be encoded as a query vector $Q \in (-1,1)^d$. To perform classification, the HD model checks the similarity between $Q$ and all class HD vectors stored in AM by the Hamming distance metric. Finally, the HD model outputs the class with the minimum distance as the prediction.

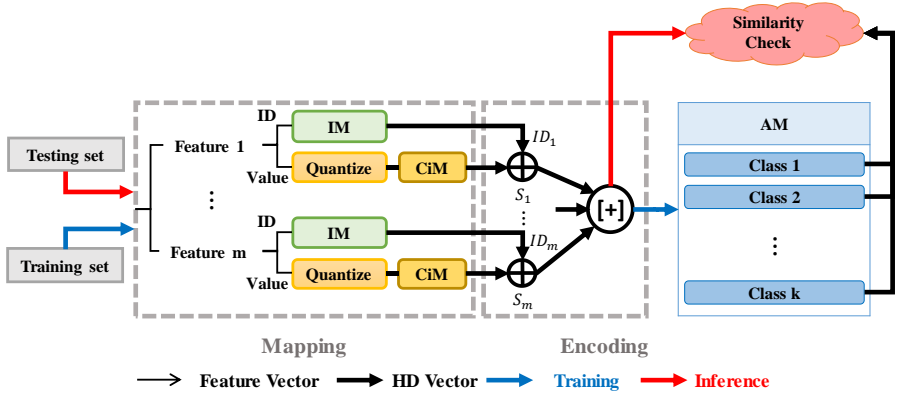

**Fig. 1** The processing flow chart of general HD computing.

## 3 Proposed Task-Projected Hyperdimensional Computing

In this section, we propose Task-projected Hyperdimensional Computing (TP-HDC) to realize multi-task learning in an HD model. We are inspired by [12], which exploits the over-parameterization in DNN to superimpose multiple models into one. This implies that only a small subspace spanned by the optimal parameters is occupied by a given task. We observe that HD computing shows a similar phenomenon, where only a small subspace spanned by class HD vectors in AM is relevant to a given task. Based on this observation, MTL can be feasible if the massive hyperspace is partitioned appropriately for each task.

### 3.1 AM Table for Multi-task Learning

Before diving into the illustration of the proposed scheme, we first introduce the definition of AM table supporting multiple tasks and its notation. Following the training flow described in Section 2, each task in task sequence $\{T_1, T_2, ..., T_s\}$ generates its own AM. A total of $s$ AM are present and form a 2-dimensional AM table, as shown in Fig. 2(a). Each column of the table comprises $k$ class HD vectors. We notate the $j^{th}$ class vector of the $i^{th}$ task as $C_i^j$. If an original HD model needs to support multiple tasks, the memory requirement of storing the AM table grows linearly with the number of tasks. For resource-constrained edge devices, the memory overhead could hinder HD computing from MTL. As a result, it is more desirable to store a compressed AM with a size that is independent of the number of tasks, as shown in Fig. 2(b).

**Baseline Method:** Considering the $j^{th}$ class in Fig. 2(a), the baseline method bundles the class HD vectors of the same class from all involved tasks. As shown in (3), $s$ HD vectors in the $j^{th}$ class $\{C_1^j, C_2^j, ..., C_s^j\}$ are bundled together and form the vector $M_j$, which is shared across all tasks. Compressed AM comprises $\{M_1, M_2, \ldots, M_k\}$, where $M_j$ represents the $j^{th}$ class HD vector used by all tasks. That is, the baseline method naïvely finds the most representative vector in the hyperspace regardless of the spatial relation between tasks.

$$M_j = [C_1^j + C_2^j + \cdots + C_s^j], \; for \; j \in \{1,2,\ldots k\} \quad (3)$$

For the baseline method, the memory overhead is $s$ times less than that of AM table. However, we discover that the baseline method causes HD vectors of different tasks to occupy overlapping subspace. This induces interference between tasks and significant accuracy degradation. In the next section, we concretize the proposed TP-HDC to efficiently realize MTL in an HD model with a lower accuracy drop.

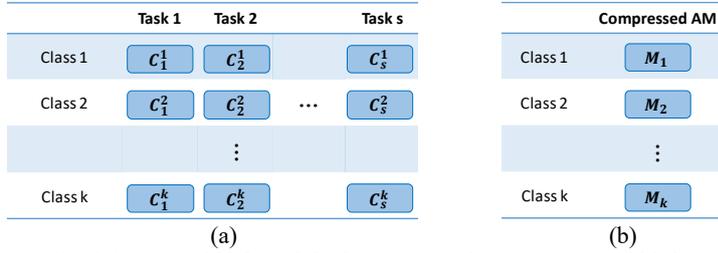

**Fig. 2** (a) AM table for original HD computing to support multiple tasks. (b) Compressed AM.

### 3.2  Orthogonalization with Task-Oriented Keys

The TP-HDC consists of the following three parts, including generation of task-oriented keys, composition with task-oriented projection, and decomposition:

**Generation of Task-Oriented Keys:** We propose to leverage the peculiar property in the Hamming space, the normalized Hamming distance from any given point in the hyperspace to a randomly drawn point highly concentrates at 0.5 [3]. Namely, two random HD vectors are approximately orthogonal (unrelated) due to hyper-dimensionality. Based on this fact, each task is assigned a task-oriented key generated at random, denoted as $\{P_1, P_2, \ldots, P_s\}$. These keys can be used for projection to achieve a division of the hyperspace in the following step of TP-HDC.

**Composition with Task-Oriented Projection:** To utilize the unused capacity of the HD model more efficiently, orthogonalization of class HD vectors of the same class, e.g., $\{C_1^j, C_2^j, \ldots, C_s^j\}$ for the $j^{th}$ class is required. With task-oriented keys generated in the previous step, we bind the keys and the class HD vectors for each task. The effect of binding projects originally close HD vectors $\{C_1^j, C_2^j, \ldots, C_s^j\}$ to different zones of the hyperspace since pseudo-randomly generated keys are approximately orthogonal. The new class HD vector ($M_j$) is formed by bundling the generated $s$ HD vectors, as shown in (4). By projecting the class HD vectors of different tasks into near-orthogonal hyperspaces, TP-HDC can mitigate the information loss caused by directly bundling class HD vectors, as implemented by the baseline method.

$$M_j = \sum_{i=1}^{s} C_i^j \oplus P_i$$
$$= [C_1^j \oplus P_1 + C_2^j \oplus P_2 + \cdots + C_s^j \oplus P_s], \; for \; j \in \{1,2,\ldots k\} \quad (4)$$

**Decomposition:** Retrieval of class HD vector of the $j^{th}$ class in $m^{th}$ task $C_m^j$, is ensured by binding $M_j$ with $P_m$, as shown in (5). The resulting vector consists of $C_m^j$ and noise $\epsilon$ because the vectors are stored in superposition. Despite the presence of $\epsilon$, TP-HDC can still be reliable because HD computing is robust against noise [3].

$$\hat{C}_j = M_j \oplus P_m = [C_1^j \oplus P_1 + C_2^j \oplus P_2 + \cdots + C_s^j \oplus P_s] \oplus P_m$$
$$= [C_1^j \oplus P_1 \oplus P_m + \cdots + C_m^j \oplus P_m \oplus P_m + \cdots + C_s^j \oplus P_s \oplus P_m]$$
$$= [C_m^j + \epsilon], \; for \; j \in \{1,2,\ldots k\} \quad (5)$$

### 3.3 Training and Inference in TP-HDC

The framework of TP-HDC is depicted in Fig. 3, and the procedure of training and inference is summarized in Algorithm 1:

**Training:** Given a task sequence, $T = \{T_1, T_2, \ldots, T_s\}$ each with $k$ classes, $s$ different AMs are updated using the general HD computing training flow described in Section 2. $C \in \mathbb{R}^{k \times s \times d}$ is a three-dimensional matrix. The first two axes of $C$ represent the number of classes and tasks, respectively, and the last axis of $C$ represents the dimensionality of HD vectors. We first generate task-oriented projection keys and denoted them as $P = \{P_1, P_2, \ldots, P_s\}$. $P$ helps achieve a division of space and project class HD vectors of different tasks to separate subspaces with equation (4). The compressed AM is $M \in \mathbb{R}^{k \times d}$, whose size is independent of the number of tasks.

**Inference:** Given a task, the HD model produces the query HD vector $Q$ in the inference phase by processing a testing sample with the same mapping and encoding modules used in the training phase. After retrieving all class HD vectors of the specific task $\hat{C} = \{\hat{C}_1, \hat{C}_2, \ldots, \hat{C}_k\}$ with equation (5), the classification result is the class in which the corresponding class HD vector has the smallest Hamming distance with $Q$, see equation (6).

$$Prediction = \underset{j}{argmin} \, Ham(\hat{C}_j, Q) \quad (6)$$

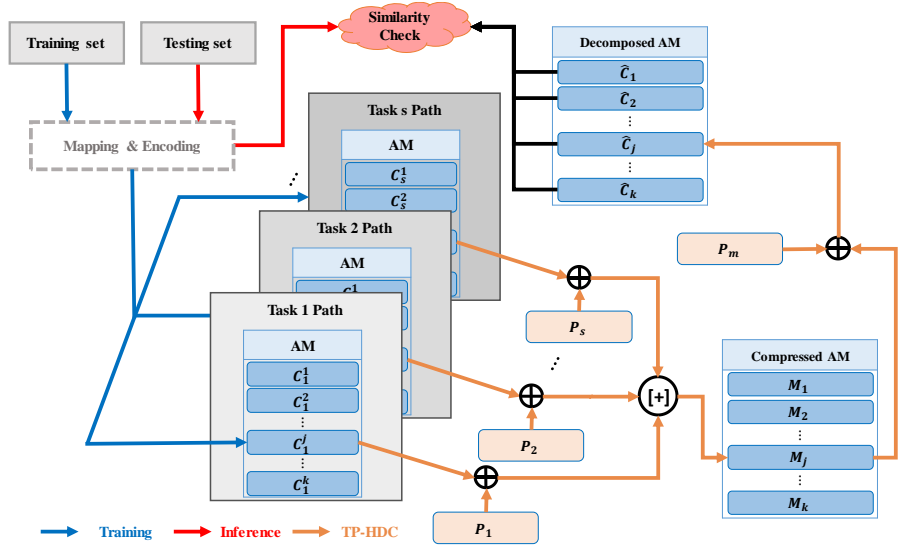

**Fig. 3** The framework of the proposed task-projected HD computing (TP-HDC).

| Algorithm 1 Task-projected HD Computing |
|---|
| **Input:**    $T = \{T_1, T_2, T_3, ..., T_s\}$ – task sequence, each with $k$ classes |
|            $mode$ – training phase or inference phase of the $m^{th}\ task$ |
|            $d$ – dimensionality of the HD model |
| **Output:** $P = \{P_1, P_2, P_3, ..., P_s\}$ – task-oriented projection keys |
|            $M$ – compressed AM |
|            $Y$ – prediction |
| 1:   **Initialize** $C \leftarrow 0_{k,s,d}$ ; $M \leftarrow 0_{k,d}$ ; $\hat{C} \leftarrow 0_{k,d}$ |
| 2:   **if** *mode* = training **do** |
| 3:     **for** each task in $T$ **do** |
| 4:       **for** each training data **do** |
| 5:         update $C$ |
| 6:     generate projection keys $P$ |
| 7:     **for** $(j = 1: k)$ **do** |
| 8:       $M[j] = \sum_{i=1}^{s} C[j][i] \oplus P_i$ |
| 9:   **if** *mode* = inference **do** |
| 10:     **for** $(j = 1: k)$ **do** |
| 11:       $\hat{C}[j] = M[j] \oplus P_m$ |
| 12:     make prediction $Y$ based on $\hat{C}$ |

# 4 Experimental Settings and Simulation Results

## 4.1 Comparisons

We compare the proposed TP-HDC with two different approaches tackling the multi-task learning problem in HD computing, including the baseline method and the ideal method.

**Baseline method:** As mentioned in Section 3.1, the baseline method naïvely finds the most representative vector among all tasks with bundling operation, causing severe interference between different tasks. Therefore, the baseline method can be regarded as the model telling us what happens if we do nothing to explicitly retain information from the previous tasks.

**Ideal Method:** The ideal benchmark considers the case where computing resources are unconstrained so that all tasks can have their own AM. Therefore, the performance of the ideal method can be viewed as the upper bound of our evaluation since the class HD vectors are stored without any information loss.

## 4.2 Dataset and Experimental Setup

We evaluate the effectiveness of our proposed TP-HDC on Split MNIST, a standard benchmark for multi-task learning [14]. Following the experiment setting of [14] with minor modifications, we split ten digits into disjoint sets. Each set corresponds to a specific task in $T = \{T_1, T_2, ..., T_s\}$, where $T_i$ aims at discriminating between $k_i$ digits $\{D_i^1, D_i^2, ..., D_i^{k_i}\}$. We fix the dimensionality of HD computing at $d = 5000$, where the performances of all HD computing models saturate. Mapping and encoding modules are shared across all tasks, meeting the expectations of MTL. Moreover, we vectorize the gray images of digits in MNIST to form 784-dimensional feature vectors and pre-process pixel values using min-max normalization. All experiments are conducted on 100 independent runs to get the final averaged simulation results.

## 4.3 Performance Analysis

First, we evaluate our proposed TP-HDC with a three-task MTL configuration. Each of the tasks, namely task A, task B, and task C contains three digits different from those of the other two tasks. HD models are sequentially trained on task A, task B, and task C. We observe that 100 training samples are enough for the convergence of all HD models in each task. Therefore, we train each task for 100 steps, and a training sample is randomly drawn to update AM in each step.

Fig. 4 illustrates the learning curve of the different methods on split MNIST. Compared with the ideal method, the baseline method suffers from catastrophic forgetting, resulting in around 20% accuracy drop on task A and task B. Furthermore, task A has occupied the subspace that task B and task C need to learn for classification, causing information loss for task B and task C and bringing about a 15% accuracy drop. On the other hand, the accuracy of the proposed TP-HDC only drops by 3.6%, 3.9%, 2.9% on task A, task B, and task C, respectively.

To validate the generalization ability of our model, we also evaluate TP-HDC on the five-task case. The experimental setup is almost the same as the three-task case except that each task contained two digits. For the baseline method, Fig. 5 shows that the five tasks tend to interfere with each other severely like that in the three-task case, leading to a 16.5% accuracy drop. In comparison, TP-HDC provides around 12.8% improvement in averaged accuracy compared with the baseline method and performs closely to the ideal benchmark consistently, with a slight 3.7% accuracy drop on average. By efficiently separating the subspaces, TP-HDC mitigates the effect of interference between tasks and improves the performance of sequential training on multiple tasks.

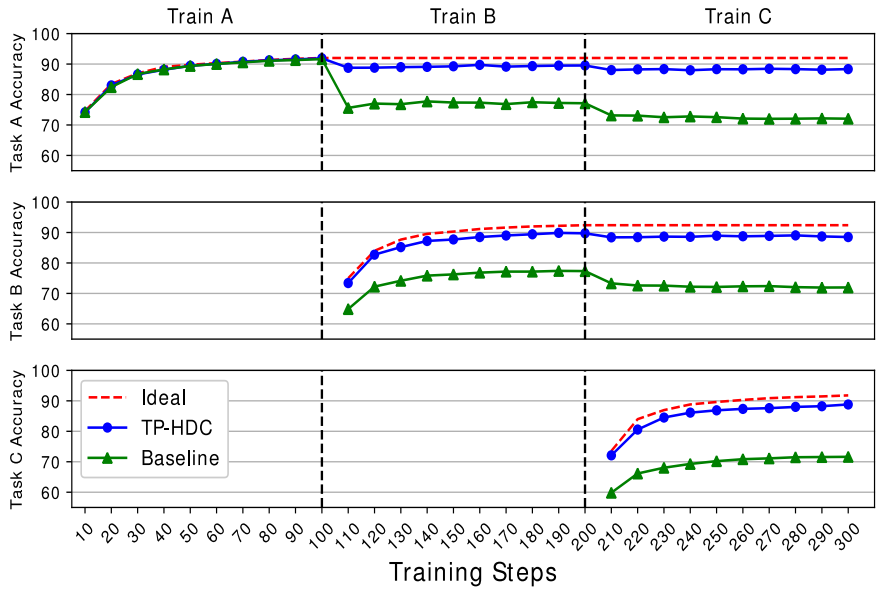

**Fig. 4** The learning curves of different methods trained on split MNIST. Tasks A, B, and C are assigned with disjoint sets of MNIST digits. The vertical dashed lines imply the transitions of the training procedure of different tasks. The top plot shows the accuracy of task A. The middle plot and bottom plot indicate the accuracy of tasks B and C, respectively.

Table 1 shows the performance of TP-HDC on split MNIST in different cases. The high standard deviation of the accuracy of the baseline method implies instability. By contrast, the results demonstrate both effectiveness ($< 4\%$ accuracy drop compared with the ideal benchmark) and stability (lower variance compared with the baseline method) of TP-HDC.

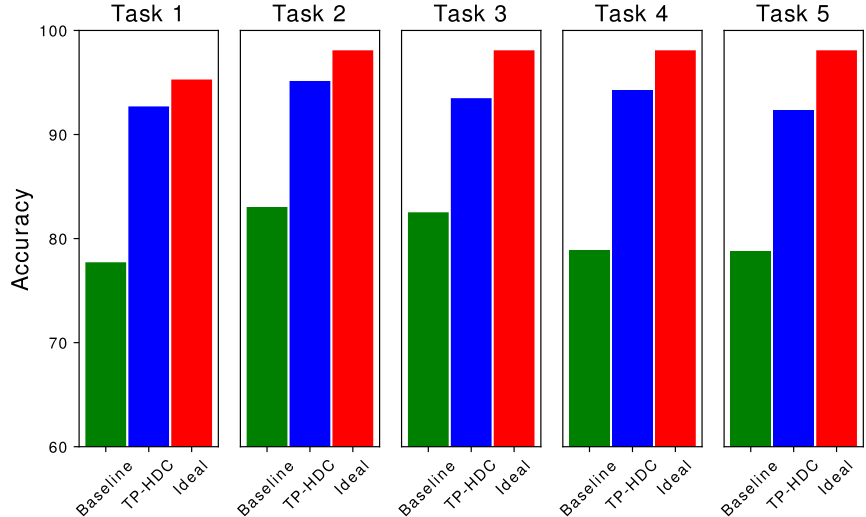

**Fig. 5** Classification accuracy of the five-task case of split MNIST. TP-HDC provides around 12.8% improvement in averaged accuracy compared with the baseline method.

**Table 1** Performance of different methods on split MNIST of different numbers of tasks.

|  | Accuracy $\pm$ Std. (%) | | | |
| --- | --- | --- | --- | --- |
| # of Tasks | 2 | 3 | 4 | 5 |
| Baseline | 88.1±8.5 | 83.6±12.2 | 79.9±13.7 | 79.2±13.1 |
| **Proposed TP-HDC** | 94.0±4.6 | 94.1±4.6 | 92.1±6.0 | 91.9±5.9 |
| Ideal case | 95.7±2.7 | 95.9±3.1 | 95.7±3.0 | 95.6±3.2 |

### 4.4 Memory Footprint Analysis

The memory requirement for the ideal method to store AM is $\mathcal{O}(s \times k)$, where $s$ and $k$ are the number of tasks and classes, respectively. Since TP-HDC just needs to store the projection keys of each task for decomposing class HD vectors in the inference phase, the memory footprint for TP-HDC only requires $\mathcal{O}(t + c)$. Moreover, instead of storing all the projection keys in the memory, linear feedback shift register (LFSR) can be utilized to generate pseudo-random patterns as a hardware-friendly approach for implementation. This makes TP-HDC more efficient for multi-task learning with negligible memory overhead.

## 5 Conclusions

To the best of our knowledge, in this paper, we first investigate the feasibility of applying multi-task learning to HD computing. To avoid catastrophic forgetting, we propose TP-HDC to exploit redundant dimensionality in the hyperspace. By separating subspaces for each task with task-oriented keys, the information loss caused by the interference between tasks is effectively reduced. Based on our experimental results, TP-HDC outperforms the baseline method on split MNIST by 12.8% accuracy on average and can be implemented with negligible memory overhead.

**Acknowledgments.** This work is supported by the Ministry of Science and Technology of Taiwan (MOST 106-2221-E-002-205-MY3 and MOST 109-2622-8-002-012-TA), National Taiwan University, and Pixart Imaging Inc.